\documentstyle[12pt]{article}
\def \beq{\begin{equation}}
\def \eeq{\end{equation}}
\def \bea{\begin{eqnarray}}
\def \eea{\end{eqnarray}}
\def \s{\sqrt{2}}

\def\fbar{\overline{f}}

\def\Kbar{\overline{K}}

\def\tl{\tan\theta_c}

\def\to{\rightarrow}
\begin{document}
\begin{titlepage}
\rightline{TECHNION-PH-2000-30}
\rightline{hep-ph/0011392}
\rightline{November 2000}
\bigskip
\bigskip

\large
\centerline {\bf CKM Phases from CP Asymmetries~\footnote
{To be published in Proceedings of Beauty 2000, Kibbutz Maagan, Israel,
September 13--18, 2000, edited by S. Erhan, Y. Rozen, and P. E. Schlein,
Nucl.\ Inst.\ Meth. A, 2001.}}
\normalsize
\vskip 2.0cm
\centerline {Michael Gronau}
\centerline {\it Department of Physics}
\centerline {\it Technion - Israel Institute of Technology, 32000 Haifa,
Israel}
\vskip 4.0cm

\centerline {\bf Abstract}
\vskip 1.0cm
Measuring phases in the Cabibbo-Kobayashi-Maskawa matrix through CP
asymmetries in $B$ decays is a major goal of current and future experiments.
Methods based on charge-conjugation and isospin symmetries involve very little
theoretical uncertainties, while schemes based on flavor SU(3) involve
uncertainties due to SU(3) breaking. Resolving these uncertainties
requires further studies involving a dialogue between theory and experiments.

\vfill
\end{titlepage}

\newpage

\section{Introduction}

Two important developments in $K$ and $B$ physics took place in the past
two years towards a better understanding of CP violation. First, {\em after
thirty five years of search} direct CP violation was finally observed in
$K^0\to \pi\pi$ \cite{K}, confirming {\em qualitatively} a prediction of the
Cabibbo-Kobayashi-Maskawa (CKM) picture \cite{eps}. A precise calculation of
the measured effect within
the CKM framework is precluded by theoretical uncertainties in evaluating
hadronic matrix elements of low energy effective weak transition operators.

Second, {\em after one year of operation} the two asymmetric $e^+e^-$
$B$-factories, the PEP-II machine at SLAC \cite{BaBar} and the KEK-B collider
in Japan \cite{Belle},
demonstrated with their new detectors, BaBar and Belle, a promising
discovery potential for CP asymmetry in $B^0\to J/\psi K_S$ .
This followed a few earlier attempts to measure this asymmetry, by the CDF
Collaboration at the Tevatron \cite{CDF} involving comparable errors,
and by the OPAL and ALEPH Collaborations at LEP \cite{LEP} involving larger
errors. The present world average value, $\sin 2\phi_1\equiv \sin 2\beta =
0.49 \pm 0.23$ \cite{JR},
extracted from all five measurements, is consistent with the CKM prediction
where the asymmetry is given by $\sin 2\beta$ \cite{BS}. At the same
time direct CP asymmetries were measured by the CLEO Collaboration at CESR
\cite{CLEO} for several self-tagging hadronic and radiative $B$ decays.
Whereas
experimental errors are somewhat smaller than in $B^0\to J/\psi K_S$,
CKM asymmetry predictions for these decays involve large theoretical
uncertainties as in the case of $K^0 \to\pi\pi$. At this point all present
direct CP measurements are consistent with zero asymmetries.

In the next few years substantial improvements in $B$ decay asymmetry
measurements are expected. Assuming the CKM framework, large nonzero
asymmetries should be observed in certain decays. Of particular importance
are decays such as $B^0\to J/\psi K_S$, where the asymmetry determines
a CKM weak phase {\em within an excellent approximation} \cite{MG}. In other
cases, to be discussed in this talk, the determination of weak phases requires
{\em measuring rates and asymmetries in several symmetry-related processes}.
Relations
between processes which permit phase determinations follow in the best
case from accurate symmetries of strong interactions, such as
charge-conjugation
or isospin. In other cases, where less accurate symmetries such as flavor
SU(3)
must be employed, one may obtain information about symmetry breaking from
other $B$ decay processes.

The purpose of this talk is to review some of the methods suggested in the
past
ten years, based on this general idea for determining the two ``problematic"
weak phases $\phi_2 \equiv \alpha$ and $\phi_3 \equiv \gamma$. Quite a few
$B$ and $B_s$ decay modes were shown to be useful in this respect. We choose
a few relatively simple examples to represent a much broader effort.

Charge-conjugation
symmetry is used in Section 2 for two measurements of $\gamma$, while the
extraction of $\alpha$ in Section 3 is based on isospin symmetry. We refer to
these methods as accurate determinations of weak phases since they are based
on good symmetries of strong interactions. In some cases these methods involve
the experimental challenge of measuring rare processes with sufficient
precision, which may not be achieved in the first round of experiments.
Various
applications of flavor SU(3) symmetry in charmless $B$ decays are discussed in
Section 4, stressing in particular the use of U-spin symmetry in determining
the weak phase $\gamma$. We argue that uncertainties due to SU(3) breaking
effects may be reduced, or even completely eliminated, by measuring these
effects in certain processes. We conclude in Section 5.

In each of these schemes one measures a simple trigonometric function of a
phase, such as the sine of twice the angle ($\sin 2\alpha$) or the square
of the sine of an angle ($\sin^2\gamma$). This leaves discrete ambiguities
in the solutions for the angles themselves. Such ambiguities, which
may hide new physics effects, can be resolved by other rather challenging
measurements of different trigonometric functions of the phases, and will
be discussed elsewhere at this conference \cite{ambig}.
In the presence of an ambiguity a conservative strategy will be to choose a
phase value consistent with the CKM framework. This should eventually
improve to a higher precision our knowledge of the CKM mixing matrix.
Alternatively, if inconsistencies are found, they would provide probes
for new physics.

Before starting a discussion of specific methods, we mention a more
ambitious approach, discussed elsewhere at this conference \cite{Neubert,Li}.
Hadronic $B$ decay amplitudes into two light mesons are calculated within QCD
to leading order in a heavy quark expansion in terms of weak phases
and several nonperturbative quantities including form factors, light cone
quark distributions in mesons, and chirality enhanced large corrections which
occur formally at order $1/m_b$. This approach is comparable to
the calculation of direct CP violation in $K^0\to \pi\pi$ \cite{eps}, with the
disadvantage that strong phases cannot be measured as in $K$ decays.
The advantage lies in the possibility, at least in principle, of carrying out
a systematic heavy quark expansion. When applied to weak phase determinations,
this approach suffers from the same uncertainties due to SU(3) breaking
as discussed in Section 4. The schemes discussed in this section for
controlling SU(3) breaking effects apply also to this approach.

\section{Accurate determinations of $\gamma$}
\subsection{$\gamma$ from $B\to D K$}

In $B^+\to D K^+$ two amplitudes interfere due to color-favored $\bar b\to\bar
c u \bar s$ and color-suppressed $\bar b\to \bar u c \bar s$ transitions.
This provides a few variants of a basically very simple idea \cite{GW} for
determining the relative weak phase $\gamma$ between the two amplitudes.
We will describe two variants, in both of which one is trying the
measure $\gamma$ through this interference \cite{other}.
Let us discuss these two cases in some detail.

\medskip
1. {\it $B$ decay to $K$ and flavor specific $D^0$ modes} \cite{ADS}
\medskip

\noindent
The three-body decay $B^+ \to (K^-\pi^+)_D K^+$, where the $K^-\pi^+$ pair has
a $D^0$ invariant mass, involves an interference between two cascade
amplitudes,
\beq
Aa_{K\pi} \equiv A(B^+\to D^0 K^+)A(D^0\to K^-\pi^+)~~,
\eeq
and
\beq
\bar A\bar a_{K\pi} \equiv A(B^+ \to \bar D^0 K^+)A(\bar D^0\to K^-\pi^+)~~.
\eeq
The first amplitude $A$,
due to $\bar b \to\bar u c \bar s$, is color-suppressed and subsequently the
$D^0$ decays into a Cabibbo-favored mode with amplitude $a_{K\pi}$. The second
amplitude $\bar A$ from $\bar b \to \bar c u \bar s$ transition is
color-favored,
and subsequently $\bar D^0$ decays with a doubly Cabibbo-suppressed amplitude
$\bar a_{K\pi}$. The
relative weak phase between $A$ and $\bar A$ is $\gamma$, their strong
phase-difference will be denoted $\delta,~{\rm Arg}(A/\bar A)=\delta +
\gamma$, and the relative phase between $a_{K\pi}$ and $\bar a_{K\pi}$
(including a relative weak phase $\pi$)
will be denoted $\Delta_{K\pi}\equiv {\rm Arg}(a_{K\pi}/\bar a_{K\pi})$.
Omitting a common phase space factor,
\bea\label{KpiK}
A(B^+ \to (K^-\pi^+)_D K^+) &=& Aa + \bar A\bar a~~,\\
\Gamma(B^+ \to (K^-\pi^+)_D K^+) &=& |Aa|^2 +|\bar A\bar a|^2 +
2|A\bar A a\bar a|\cos(\delta + \Delta + \gamma)~~,
\nonumber
\eea
where $a\equiv a_{K\pi},~\bar a\equiv \bar a_{K\pi},~\Delta\equiv
\Delta_{K\pi}$.

The rate for the charge-conjugate process, $B^- \to (K^+\pi^-)_D K^-$, has a
similar expression in which $\gamma$ occurs with an opposite sign, while
strong
phases are invariant under charge-conjugation. The CP asymmetry in this
process,
involving an interference of $Aa_{K\pi}$ and $\bar A\bar a_{K\pi}$, is
proportional to $\sin(\delta + \Delta_{K\pi})\sin\gamma$.

Let us summarize the present updated information on the parameters appearing
in Eqs.~(\ref{KpiK}). The three amplitudes $\bar A,~a_{K\pi}$ and $\bar
a_{K\pi}$ have already been measured \cite{PDG, CLEODCS}. The measured ratio
$|\bar a_{K\pi}/a_{K\pi}| =
(1.21\pm 0.13)\tan^2 \theta_c = 0.062 \pm 0.007$ is consistent at 90$\%$
confidence level with flavor SU(3) symmetry, which
predicts a value of $-\tan^2\theta_c$ for the ratio of amplitudes \cite{KTWZ}.
The amplitude $A$ can be estimated as follows.
It involves a CKM factor of $|V^*_{ub}V_{cs}|/|V^*_{cb}V_{us}|\approx 0.4$
relative to $\bar A$, and is expected to be color-suppressed relative to this
amplitude by a factor of about 0.25, measured in $B\to \bar D\pi$ decays
\cite{BHP}. Thus one estimates $|A/\bar A| \sim 0.1$.

Therefore, the two amplitudes interfering in
Eqs.~(\ref{KpiK}) are anticipated to be comparable in magnitude,
$|\bar A\bar a_{K\pi}/Aa_{K\pi}|\sim 0.6$. This, and large final state phases
measured in
Cabibbo-favored $D\to K\pi$ decays \cite{CLEOD}, raised the hope \cite{ADS}
for
a possible large CP asymmetry in this process. We note, however, that the
relevant phase $\Delta_{K\pi}$ between $a_{K\pi}$ and $\bar a_{K\pi}$ vanishes
in the SU(3) limit \cite{KTWZ} and, as mentioned above, SU(3) does not seem
to be strongly broken in $|\bar a_{K\pi}/a_{K\pi}|$. A recent study \cite{UD}
suggests that $\Delta_{K\pi}$ is unlikely to be larger than about $20^\circ$.

The rate expression (\ref{KpiK}) and its charge-conjugate provide two equations
for the three unknowns: $A,~\delta + \Delta_{K\pi}$ and $\gamma$. To solve for
$\gamma$ requires observing another doubly Cabibbo-suppressed $D^0$ decay mode.
Such a study is currently under way in the $K^+ \pi^- \pi^0$ channel
\cite{smith}, and should soon provide a result for the doubly
Cabibbo-suppressed
amplitudes $\bar a_{K^+\rho^-},~\bar a_{K^{*+}\pi^-}$ and $\bar
a_{K^{*0}\pi^0}$. Assuming, for instance, a knowledge of $\bar a_{K\rho}$, two
equations identical to Eqs.~(\ref{KpiK}) and its charge-conjugate can be
written for $\Gamma(B^+ \to (K^-\rho^+)_D K^+)$ and $\Gamma(B^- \to
(K^+\rho^-)_D K^-)$ involving $a\equiv a_{K\rho},~\bar a\equiv \bar a_{K\rho},~
\Delta\equiv \Delta_{K\rho}$. This introduces in the four equations only one
new unmeasurable quantity, $\delta + \Delta_{K\rho}$, such that these equations
can be solved for $\gamma$ modulo some discrete ambiguities. The ambiguities
may be reduced by including information from other doubly Cabibbo-suppressed
modes \cite{ADS2}.

This method requires a large number of $B$'s, at least of order $10^8-10^9$.
This is obvious, since, for instance, ${\cal B}(B^+\to \bar D^0 K^+){\cal
B}(\bar D^0\to K^-\pi^+) = (4.2 \pm 1.4)\times 10^{-8}$ \cite{PDG,CLEODCS}.

\medskip
~~2. {\it $B$ decay to $K$ and $D^0$ CP-eigenstate modes} \cite{GDK}
\medskip

Neglecting very small CP violation in $D^0-\bar D^0$ mixing, one can write
neutral $D$ meson even/odd CP states (decaying, for instance, to
$K^+K^-$ or $K_S\pi^0$) as $D^0_{\pm}=(D^0 \pm \bar D^0)/\s$. Consequently,
one has up to an overall phase
\beq
\sqrt{2}A(B^+\to D^0_{\pm} K^+) = \pm |\bar A| + |A|\exp[i(\delta + \gamma)]~~.
\eeq
Let us define charge-averaged ratios of rates for positive and negative CP
states relative to rates corresponding to color-favored neutral $D$ flavor
states
\beq\label{Rpm}
R_{\pm} \equiv \frac{2[\Gamma(B^+ \to D_{\pm} K^+) + \Gamma(B^- \to D_{\pm}
K^-)]}{\Gamma(B^+ \to \bar D^0 K^+) + \Gamma(B^- \to D^0 K^-)}~~,
\eeq
and two corresponding pseudo-asymmetries
\beq\label{Apm}
{\cal A}_{\pm} \equiv \frac{\Gamma(B^+ \to D_{\pm} K^+) - \Gamma(B^-
\to D_{\pm} K^-)}
{\Gamma(B^+ \to \bar D^0 K^+) + \Gamma(B^- \to D^0 K^-)}~~.
\eeq
These quantities do not require measuring the color-suppressed rate
$\Gamma(B^+ \to D^0 K^+)$ and its charge-conjugate. One finds
\bea\label{RA}
R_{\pm} &=& 1 + |A/\bar A|^2 \pm 2|A/\bar A|\cos\delta\cos\gamma~~,\nonumber \\
{\cal A}_- &=& -{\cal A}_+ = |A/\bar A| \sin\delta \sin\gamma~~.
\eea

In principle, Eqs.~(\ref{RA}) provide sufficient information to determine the
three parameters $|A/\bar A|, \delta$ and $\gamma$, up to certain discrete
ambiguities. However, as explained above, one expects $|A/\bar A| \sim 0.1$.
Such a value would be too small to be measured with high precision from the
tiny deviation from unity of $(R_+  + R_-)/2 = 1 + |A/\bar A|^2$ .

Nevertheless, one obtains two interesting bounds
\beq\label{LIMIT}
\sin^2\gamma \leq R_{\pm}~~,
\eeq
which could potentially imply new constraints on $\gamma$ in future
experiments.
Assuming, for instance, $|A/\bar A|=0.1,~\delta=0,~\gamma=40^\circ$, one finds
$R_-=0.85$.
With $10^8~~B^+B^-$ pairs, using measured $B$ and $D$ decay branching
ratios \cite{PDG}, one estimates an error \cite{GDK} $R_-=0.85 \pm 0.05$. In
this case, Eq.(\ref{LIMIT}) excludes the range $73^\circ
<\gamma < 107^\circ$ with 90$\%$ confidence level. Including measurements
of the
CP asymmetries ${\cal A}_{\pm}$ could further constrain $\gamma$. Again, this
method would require at least $10^8-10^9$ $B$'s similar to the previous scheme.
In fact, both methods could and should be combined to improve precision
\cite{soffer}.

The large number $B$'s needed to measure an asymmetry reflects the small
color-suppressed rate $\Gamma(B^+\to D^0 K^+) \propto |A|^2$ and the combined
$D^0$ branching ratio into CP-eigenstates which is a few percent. Although
$|A|^2$ does not have to be measured, one can show in general \cite{EGR} that,
whenever an asymmetry has to be measured due to an interference between two
processes, the required number of events is dictated by the branching ratio of
the rarer process and is independent of the more frequent process. For this
reason it would be much preferable to use $B\to D_{\pm}K$ rather than
$B\to D_{\pm}\pi$.

Very recently a variant of this scheme was proposed \cite{LSS}, in which one
measures
in flavor tagged $B^0$ decays to two vector mesons, $B^0\to D^{*0}K^{*0}$,
both
the time dependence in this process and the angular dependence in
$D^{*0}\to D^0\pi^0$ and $K^{*0}\to K_S\pi^0$. Measuring interference terms
between different helicity amplitudes permits a determination of
$\sin^2(2\beta +\gamma$), involving the sum of the weak phase in $B^0-\bar
B^0$
mixing and the phase in $B$ decay. This would provide information on $\gamma$,
assuming that by the time of this measurement $\beta$ will have been
determined.
It is claimed that the sensitivity of this method is not limited by the small
color-suppressed rates as it is in the above two schemes for measuring
$\gamma$.
However, flavor tagging suppresses the number of events, and a detailed
angular
analysis may be statistics limited.

\subsection{$\gamma$ from $B_s(t)\to D_s K$}

Time-dependence in $B_s(t)\to D^-_s K^+$ is expected to
exhibit an oscillating behavior including interference of
two amplitudes, $A_s=A(B_s\to D^-_s K^+)$ and $\bar A_s=
A(\bar B_s\to D^-_s K^+)$, from quark transitions
$\bar b \to \bar c u \bar s$ and $b\to u\bar c s$, respectively. Both
amplitudes, of order $\lambda^3$, are color-allowed and involve a relative
weak phase $\gamma$. This interference
leads to a $\sin(\Delta m_s t)$ term  in the rate, as in decays to
CP-eigenstates.
For simplicity, let us first neglect the width-difference between
the two strange $B$ meson mass-eigenstates. Denoting the relative strong phase
between $A_s$ and $\bar A_s$ by $\delta_s$, one obtains the well-known result
for the general time-dependence of neutral $B$ decays \cite{tdep}
\bea\label{DsK}
\Gamma(B_s(t) \to D^-_sK^+)&=&e^{-\Gamma_st}[|A_s|^2\cos^2(\frac
{\Delta m_st}{2}) + |\bar A_s|^2\sin^2(\frac{\Delta m_st}{2})\nonumber\\
&+&|A_s\bar A_s|\sin(\delta_s+\gamma) \sin(\Delta m_s t)]~.
\eea

Due to the invariance of strong phases under charge-conjugation, the same
strong phase $\delta_s$ occurs also in decay rates for charge-conjugate
initial
and final states, $\bar B_s(t) \to D^-_sK^+,~B_s(t) \to D^+_sK^-,~\bar
B_s(t) \to D^+_sK^-$. The weak phase changes sign under charge-conjugation.
Thus, all four time-dependent rates can be expressed in terms of four
quantities, $|A_s|,~|\bar A_s|, \sin(\delta_s + \gamma)$ and $\sin(\delta_s -
\gamma)$. Measuring the time-dependence of these four processes, all
of which require flavor tagging of the initial strange $B$ meson, permits a
determination of $\gamma$ up to a discrete ambiguity~~\cite{ADK}.

A more precise expression than (\ref{DsK}) includes a dependence on the
width-difference $(\Gamma_L - \Gamma_H)/\Gamma_{\rm ave}$, which is expected
to be of order 10$-$20 $\%$. Assuming that the two exponential decays due to
two different lifetimes can be seperated by this measurement, one obtains
useful
information also from untagged rates \cite{Dunietz}:
\bea\label{two-rates}
& &\Gamma(B_s(t) \to D^-_sK^+) + \Gamma(\bar B_s(t) \to D^-_sK^+) = \\
\nonumber
& &\frac{1}{2}(|A_s|^2 + |\bar A_s|^2) (e^{-\Gamma_Lt} + e^{-\Gamma_Ht})
+ |A_s\bar A_s|\cos(\delta_s+\gamma) (e^{-\Gamma_Lt} - e^{-\Gamma_Ht})~.
\eea
The untagged decay rate into $D^+_sK^-$ has a similar expression, in which
$\cos(\delta_s+\gamma)$ is replaced by $\cos(\delta_s-\gamma)$. In order to
extract both $\cos(\delta_s+\gamma)$ and $\cos(\delta_s-\gamma)$, thus
eliminating part of the discrete ambiguity in $\gamma$, one needs
independent information about $|A_s|^2$. This information can be obtained from
the flavor tagged rates, or by relating $|A_s|^2$ through factorization
to the measured value of the CKM-favored rate $|A(B_s\to D^-_s\pi^+)|^2$.

\section{$\alpha$ from $B\to\pi\pi$}

In the CKM framework $\alpha = \pi-\beta-\gamma$. This phase occurs in the
time-dependent rate of $B^0(t)\to \pi^+\pi^-$ and would dominate its asymmetry
if only one amplitude (``tree" $T$)  contributes. In reality
this process involves a second amplitude ($P$) due
to penguin operators which carries a different weak
phase than the dominant tree amplitude.
This leads to a more general form of the time-dependent asymmetry,
which includes in addition to the $\sin(\Delta mt)$ term a $\cos(\Delta mt)$
term due to direct CP violation  \cite{MG}
\beq\label{asymmet}
{\cal A}(t) = a_{\rm dir}\cos(\Delta mt) + \sqrt{1-a^2_{\rm dir}}
\sin 2(\alpha + \theta)\sin(\Delta mt)~.
\eeq

Both $a_{\rm dir}$ and $\theta$ are given roughly by the ratio of penguin to
tree amplitudes, $a_{\rm dir}\sim 2|P/T|\sin\delta_{\pi\pi},~
\theta\sim |P/T|\cos\delta_{\pi\pi}$, where $\delta_{\pi\pi}$ is an
unknown strong phase. This measurement provides two equations for $|P/T|,~
\delta_{\pi\pi}$ and $\alpha$, which is insufficient for measuring $\alpha$.
A rough estimate of $|P/T|$, based on CKM and QCD factors, yielded some time
ago the value 0.1 \cite{Pen93}.
When flavor SU(3) is applied to relate penguin and tree amplitudes in
measured $B\to \pi\pi$ and $B\to K\pi$ \cite{pipi} one finds \cite{DGR}
$|P/T| =0.3 \pm 0.1$. As mentioned, precise knowledge of this ratio could
provide very useful information about $\alpha$ \cite{MG, P/T}. Calculations
of $|P/T|$ \cite{Neubert,AD}
involve systematic theoretical errors which are uncontrollable at present.

One way of eliminating the penguin effect is by measuring also the
time-integrated rates of $B^0\to\pi^0\pi^0$, $B^+\to\pi^+\pi^0$
and their charge-conjugates \cite{GRLO}. The three $B\to\pi\pi$ amplitudes
obey an isospin triangle relation,
\beq\label{iso}
A(B^0\to\pi^+\pi^-)/\s + A(B^0\to\pi^0\pi^0) = A(B^+\to \pi^+\pi^0)~,
\eeq
and a similar relation holds for the charge-conjugate processes.
One uses the different isospin properties
of the penguin ($\Delta I=1/2$) and tree ($\Delta I=1/2, 3/2$) contributions
and the well-defined weak phase ($\gamma$) of the tree amplitude. By
constructing the two isospin triangles one may measure the correction
to $\sin2\alpha$ in the second term of the asymmetry in Eq.~(\ref{asymmet}).

An electroweak penguin contribution could spoil this method
\cite{DH} since it involves a $\Delta I=3/2$ component.
This implies that the amplitudes of $B^+\to\pi^+\pi^0$ and its charge-conjugate
differ in phase, which introduces a correction at the level of a few
percent in the isospin analysis. However, even this
small correction can be taken into account analytically in the isospin
analysis \cite{GPY}. Other corrections, from
isospin breaking in $\pi^0-\eta$ mixing \cite{gardner}, turn out to
be small for large values of $|P/T|$.

The difficulty of measuring $\alpha$ without knowing precisely $|P/T|$
seems to be experimental rather than theoretical.
The average branching ratios obtained from three experiments \cite{pipi}
${\cal B}(B^0\to\pi^+\pi^-)=(5.6\pm 1.3)\times 10^{-6},~{\cal B}(B^+\to\pi^+
\pi^0)=(4.6\pm 2.0)\times 10^{-6}$, are somewhat lower than anticipated some
time ago \cite{BSW}. The branching ratio into two neutral pions is likely to
be smaller, since it obtains only contributions from penguin and
color-suppressed amplitudes. In the most optimistic case, when
these contributions interfere constructively, this branching ratio could lie
just below the branching ratios measured for charged pions. A small
$B^0\to\pi^0\pi^0$ branching ratio and an experimentally indistinguishable
background may cause serious difficulties in applying this method.
One will have to wait a while before measuring this branching ratio with
sufficient precision. Stringent limits on this branching ratio would impose
an interesting bound on the uncertainty in $\sin(2\alpha)$ obtained from
the asymmetry in $B^0(t)\to\pi^+\pi^-$ \cite{GQ}
\beq
\sin(\delta\alpha)\le
\sqrt{\frac{{\cal B}(B\to \pi^0\pi^0)}{{\cal B}(B^\pm\to \pi^{\pm}\pi^0)}}
\eeq
Other ways of constraining this uncertainty were discussed in \cite{charles}.

The isospin method for resolving penguin pollution in $\sin 2\alpha$ can also
be applied to $B\to \rho\pi$ decays \cite{isorhopi}, of which there exist five
charge modes some of which have already been measured \cite{pipi}, or to
$B\to a_0(980)\pi$ \cite{DK}. Studying
Dalitz plots of $B\to 3\pi$ \cite{rhopi},
in which amplitudes describing different resonance bands involve unknown
relative phases and interfere with an unknown three pion nonresonant
amplitude, may be quite challenging \cite{BaBarbook}.

\section{$\gamma$ from $B\to PP$}
\subsection{Flavor SU(3) relates $B/B_s\to\pi\pi,K\pi,K\bar K$}

A large number of charmless $B$ and $B_s$ decays to two light pseudoscalars
can be related to each other under approximate flavor SU(3) symmetry. This
program started quite a few years ago \cite{zepp} as a way of classifying
hadronic weak amplitudes in terms of quark diagrams,
and has been applied extensively for the past seven years to the
$\Delta B=1,~\Delta C=0$ low energy effective Hamiltonian for the specific
purpose of determining weak phases. Whereas the first few attempts neglected
second order electroweak penguin (EWP) contributions \cite{GHLR,GRL}, a large
variety of
proposals \cite{GPY,BKpi,NRPL,NRPRL} were made after noting \cite{DHF} the
importance of
these terms. Here we will review briefly the most common
features of these proposals, focusing on one particular result.

The low energy effective weak Hamiltonian describing $\Delta S =1$ charmless
$B$ decays, such as $B\to K\pi$, is
\cite{BBL}
\beq\label{Heff}
{\cal H}^{(s)}_{\rm eff} = \frac{G_F}{\s}\left[V^*_{ub}V_{us}\left(\sum^2_1
c_i
Q^{us}_i +\sum^{10}_3 c_i Q^s_i\right ) + V^*_{cb}V_{cs}\left(\sum^2_1 c_i
Q^{cs}_i +\sum^{10}_3 c_i Q^s_i\right )\right]~,
\eeq
where $c_i$ are scale-dependent Wilson coefficients and the flavor structure
of the various four-quark operators is $Q^{qs}_{1,2}\sim\bar b
q\bar q s,~Q^s_{3,..,6}\sim \bar b s \sum \bar q' q',~Q^s_{7,..,10}\sim
\bar b s\sum e_{q'}\bar q' q'~(q'=u,d,s,c$). In the $\Delta S =0$ Hamiltonian
${\cal H}^{(d)}_{\rm eff}$ describing $B\to \pi\pi$ one must replace
$s\rightarrow d$.
The ten operators consist of two $(V-A)(V-A)$ current-current operators
($Q_{1,2}$), four QCD penguin operators ($Q_{3,4,5,6}$), and four EWP
operators
($Q_{7,8,9,10}$) with different chiral structures.
One makes use of their following two properties:
\begin{itemize}
\item All four-quark operators, $(\bar bq_1)(\bar q_2 q_3)$, can be decomposed
into a sum of ${\overline {\bf 15}}$, ${\bf 6}$ and ${\overline {\bf 3}}$
representations \cite{zepp}. QCD penguin operators are pure
${\overline {\bf 3}}$.
\item The EWP operators with dominant Wilson coefficients, $Q_9$ and $Q_{10}$,
have a $(V-A)(V-A)$ structure, and their components transforming as given
SU(3) representations are proportional to the corresponding components of the
current-current operators
\cite{ewpVP}.
\end{itemize}

All $B/B_s\to PP$ decays (where final states belong to ${\bf 1, 8}$ and
${\bf 27}$) can then be expressed in terms of
five SU(3) reduced amplitudes, or alternatively in terms of five independent
combinations of eight diagrams \cite{GHLR}: Tree ($T$), Color-suppressed ($C$),
Penguin ($P$), Annihilation ($A$), Exchange ($E$), Penguin Annihilation
($PA$),
EWP ($P_{EW}$) and Color-suppressed EWP ($P^c_{EW}$).
A useful proportionality relation between EWP and current-current (``tree")
operators is
\beq\label{15}
{\cal H}^{(q)}_{EWP}(\overline{\bf 15}) = -\frac32 \frac{c_9+c_{10}}{c_1+c_2}
\frac{V^*_{tb}V_{tq}}{V^*_{ub}V_{uq}}
{\cal H}^{(q)}_T(\overline{\bf 15})~,~~~~q=s,d~~,
\eeq
where $(c_9+c_{10})/(c_1+c_2)  \approx -1.12\alpha$. This relation between
EWP and tree amplitudes simplifies the analysis in certain cases.

Consider, for instance, $B\to (K\pi)_{I=3/2}$, where
$|I=3/2\rangle=|K^0\pi^+\rangle + \s|K^+\pi^0\rangle$. This process obtains
only contributions from $\overline{\bf 15}$ EWP and tree
operators, which are proportinal to each other and involve a
common strong phase. Their ratio is given by
\beq\label{del}
-\delta_{EW}\ e^{-i\gamma} = -\frac32 \frac{c_9+c_{10}}{c_1+c_2}\frac{V^*_{tb}
V_{ts}}
{V^*_{ub}V_{us}}=-(0.65\pm 0.15)\ e^{-i\gamma}~.
\eeq
This feature can be used to obtain a bound on $\gamma$ in $B^{\pm}\to K\pi$
\cite{NRPL}. The result can be summarized as follows.
Defining a charge-averaged ratio of rates
\beq\label{R*}
R^{-1}_*\equiv \frac{2[B(B^+\to K^+\pi^0) + B( B^-\to K^-\pi^0)]}
{B(B^+\to K^0\pi^+) + B(B^-\to \bar K^0\pi^-)}~,
\eeq
one derives the following inequality, to leading order in small quantities
\beq\label{bound}
|\cos\gamma - \delta_{EW}| \ge \frac{|1-R^{-1}_*|}{2\epsilon}~,
\eeq
where \cite{pipi,GRL}
\beq\label{eps}
\epsilon =\frac{|V^*_{ub}V_{us}|}{|V^*_{tb}V_{ts}|}\frac{|T+C|}{|P+EW|}=
\sqrt2 \frac{V_{us}}{V_{ud}}\frac{f_K}{f_\pi}
\frac{|A(B^+\to \pi^0\pi^+)|}{|A(B^+\to K^0\pi^+)|} = 0.20\pm 0.05~.
\eeq
In the above isospin symmetry relates the dominant Penguin
amplitudes in $B^+\to K^+\pi^0$ and $B^+\to K^0\pi^+$. SU(3) is used in
the ratio (\ref{del}) of subdominant EWP and tree amplitudes, and SU(3)
breaking is introduced through $f_K/f_{\pi}$ in (\ref{eps}) when evaluating
the tree amplitude in $B^+\to K^+\pi^0$.

A useful constraint on $\gamma$ follows for $R^{-1}_*\ne 1$. The error of
the present average value \cite{pipi}, $R^{-1}_*=1.45\pm 0.46$, ought to be
reduced before drawing firm conclusions.
Further information about $\gamma$, applying also to the case $R^{-1}_{*}=1$,
can be obtained by measuring separately $B^+$ and $B^-$ decay rates
\cite{NRPRL}.
The solution obtained for $\gamma$ involves uncertainties due to SU(3) breaking
in subdominant amplitudes and an uncertainty in $|V_{ub}/V_{cb}|$, both of
which
affect the value of $\delta_{EW}$. Combined with errors in
$\epsilon \propto |A(B^+\to\pi^+\pi^0)/A(B^+\to K^0\pi^+)|$, and in
rescattering effects to be discussed below, the resulting uncertainty in
$\gamma$ is unlikely to be smaller than 10 or 20 degrees \cite{NRPRL}.

\subsection{U-spin in charmless $B$ decays}

A subgroup of flavor SU(3), discrete U-spin symmetry interchanging $d$ and $s$
quarks, plays a
particularly interesting and quite general role in charmless $B$ decays
\cite{Uspin}. Consider the effective Hamiltonian in  Eq.~(\ref{Heff}).
Each of the four-quark operators represents an $s$
component (``down") of a U-spin doublet, so that one can write in short
\beq\label{Us}
{\cal H}^{(s)}_{\rm eff} = V^*_{ub}V_{us}U^s + V^*_{cb}V_{cs}C^s~,
\eeq
where $U$ and $C$ are U-spin doublet operators. Similarly, the effective
Hamiltonian responsible for $\Delta S =0$ decays involves $d$ components
(``up" in U-spin) of corresponding operators multiplying CKM factors
$V^*_{ub}V_{ud}$ and $V^*_{cb}V_{cd}$,
\beq\label{Ud}
{\cal H}^{(d)}_{\rm eff} = V^*_{ub}V_{ud}U^d + V^*_{cb}V_{cd}C^d~.
\eeq

The strucure of the Hamiltonian implies a general relation between
two decay processes, $\Delta S=1$ and  $\Delta S =0$, in which initial and
final states are obtained from each other by a U-spin transformation,
$U: d\leftrightarrow s$. Writing the $\Delta S=1$ amplitude as
\beq\label{s}
A(B\to f,~\Delta S =1) = V^*_{ub}V_{us}A_u + V^*_{cb}V_{cs}A_c~,
\eeq
the corresponding $\Delta S =0$ amplitude is given by
\beq\label{d}
A(UB\to Uf,~\Delta S =0) = V^*_{ub}V_{ud}A_u + V^*_{cb}V_{cd}A_c~.
\eeq
Here $A_u$ and $A_c$ are complex amplitudes involving CP-conserving phases.
The amplitudes of the  corresponding charge-conjugate processes are
\beq\label{sbar}
A(\bar B\to \fbar,~\Delta S =-1) = V_{ub}V^*_{us}A_u + V_{cb}V^*_{cs}A_c~,
\eeq
and
\beq\label{dbar}
A(U\bar B\to U\fbar,~\Delta S =0) = V_{ub}V^*_{ud}A_u + V_{cb}V^*_{cd}A_c~.
\eeq
Unitarity of the CKM matrix \cite{Jarl}, ${\rm
Im}(V^*_{ub}V_{us}V_{cb}V^*_{cs})
= - {\rm Im}(V^*_{ub}V_{ud}V_{cb}V^*_{cd})$,
implies, denoting CP rate differences by $\Delta$
\bea\label{asym}
\Delta(B \to f) & \equiv & \Gamma(B \to f) - \Gamma(\bar B\to \fbar) =
\nonumber
 \\
-\Delta(UB \to Uf) & \equiv & - [\Gamma(UB \to Uf) - \Gamma(U\bar B \to
U\fbar)]~.
\eea

Namely, {\em CP rate differences in decays which go into one another under
interchanging $s$ and $d$ quarks have equal magnitudes and opposite signs}.
This rather powerful result, following from U-spin within the CKM framework,
can be demonstrated in numerous decay processes, including two body, quasi-two
body, and multibody hadronic and radiative $B$ decays. In view of this result,
it is rather easy to
look for physics beyond the standard model in pairs of U-spin related
processes. Since it seems unlikely that strong phases change sign under U-spin
breaking, measuring asymmetries with equal signs in such a pair
would be a clear signal for new physics.

Let us focus our attention on six U-spin related pairs of processes (out of
a total of sixteen decays) of the type $B,B_s\to\pi\pi,~K\pi,~K\Kbar$:
\bea
1.~B^0 \to K^+\pi^-~~vs.~~B_s \to \pi^+ K^-~,
~~&2.~B_s \to K^+ K^-&vs.~~B^0 \to \pi^+\pi^-~,\nonumber \\
3.~B^0 \to K^0\pi^0~~vs.~~~~B_s \to \Kbar^0\pi^0~,
~~&4.~B^+ \to K^0 \pi^+&vs.~~B^+ \to \Kbar^0 K^+~,\nonumber \\
5.~B_s \to K^0 \Kbar^0~~vs.~~B^0 \to \Kbar^0 K^0~,
~~&6.~B_s \to\pi^+\pi^-&vs.~~B^0 \to K^+ K^-~.\nonumber
\eea
Equalities of CP rate-differences within each of these pairs can be used to
test the validity of U-spin symmetry.

The first five pairs of processes are dominated by a large penguin
amplitude $P$,
such that the corresponding branching ratios are of order $10^{-5}$. The
amplitudes of the last
pair involve the combination $PA + E$, which is expected to be
very small \cite{GHLR} unless amplified by rescattering \cite{rescat}.
Neglecting rescattering, one estimates ${\cal B}(B^0 \to K^+K^-) \sim
{\cal O}(10^{-7}-10^{-8})$ \cite{Li}. To reach this level, the present
experimental upper limit \cite{pipi}, ${\cal B}(B^0\to K^+K^-) <
1.9\times 10^{-6}$, should be improved by one or two orders of magnitude.
Assuming that $PA + E$ can be neglected relative to $P$, one has
\beq\label{SU3}
A(B_s \to K^+ K^-) \approx  A(B^0 \to K^+\pi^-)~,~~
A(B_s \to \pi^+ K^-) \approx  A(B^0 \to \pi^+\pi^-)~.
\eeq
In the approximation of factorized hadronic amplitudes \cite{Neubert,Li},
U-spin breaking is introduced through the ratio of corresponding form factors,
\bea\label{su3}
& &A(B_s\to K^+ K^-)/A(B^0\to K^+\pi^-) =
A(B_s\to K^-\pi^+)/A(B^0\to \pi^+\pi^-) \nonumber \\
& &= F_{B_s K}(m^2_K)/F_{B\pi}(m^2_K)\approx
F_{B_s K}(m^2_{\pi})/F_{B\pi}(m^2_{\pi})~.
\eea
The approximate equality in ratios of form factors holds to within $1\%$.
{\it The rates of these four processes can be used not only to determine the
U-spin breaking factor in the ratio of amplitudes, but also to check the
factorization assumption by finding equal ratios of amplitudes in the two
cases.}

\subsection{$\gamma$ from $B/B_s\to K\pi$}

The processes in (\ref{SU3}) play a useful role in determining
$\gamma$. We describe a scheme based on $K\pi$
decays of $B^0$ and $B_s$ mesons \cite{bskpi},  complementary to
studying time-dependence in $B^0(t)\to \pi^+\pi^-$ and
$B_s(t)\to K^+ K^-$ \cite{flei}.

Writing the amplitudes for $B^0\to K^+\pi^-$ and $B_s\to K^-\pi^+$ as in
Eqs.~(\ref{s}) and (\ref{d}), respectively, we note that the rates for these
processes and their charge-conjugates depend on four quantities,
$\vert V^*_{ub}V_{us}A_u\vert,~\vert V^*_{cb}V_{cs} A_c\vert,~\delta_{K\pi}
\equiv {\rm Arg}(A_u A^*_c)$ and $\gamma\equiv {\rm Arg}(-V^*_{ub}V_{ud}V_{cb}
V^*_{cd})$. Because of the equality of CP rate-differences in the two
processes, a determination of $\gamma$ requires another input. This input
is provided by $|A(B^+\to K^0\pi^+)|=|V^*_{cb}V_{cs} A_c|$, where small
rescattering corrections are neglected as argued above.

Defining two charge-averaged ratios of rates
\beq
R \equiv \frac{\Gamma(B^0 \to K^{\pm} \pi^{\mp})}
{\Gamma(B^\pm \to K \pi^{\pm})}~,~~~
R_s \equiv \frac{\Gamma(B_s \to K^{\pm} \pi^{\mp})}
{\Gamma(B^{\pm} \to K \pi^{\pm})}~~,
\eeq
and CP violating pseudo-asymmetries
\beq
{\cal A}_0 \equiv \frac{\Delta(B^0 \to K^+ \pi^-)}
{\Gamma(B^{\pm} \to K \pi^{\pm})}~,~~
{\cal A}_s \equiv \frac{\Delta(B_s \to K^- \pi^+)}
{\Gamma(B^{\pm} \to K \pi^{\pm})}~~,
\eeq
one finds
\beq \label{eqn:R}
R = 1 + r^2 + 2 r \cos \delta_{K\pi} \cos \gamma~~~,
\eeq
\beq \label{eqn:Rs}
R_s = \tl^2 + (r/\tl)^2 - 2 r \cos \delta_{K\pi} \cos \gamma~~~,
\eeq
\beq \label{A0s}
A_0 = - A_s = -2r \sin \delta_{K\pi}\sin \gamma~~,
\eeq
where $r\equiv \vert V^*_{ub}V_{us}A_u\vert/\vert V^*_{cb}V_{cs} A_c\vert$.
SU(3) breaking can be checked in (\ref{A0s}) and used for improving the
precision in $\gamma$ obtained from these four quantities.
It is estimated \cite{bskpi}
that a precision of $10^\circ$ in $\gamma$ can be achieved in experiments
to be performed at the Fermilab Tevatron Run II program \cite{RUNII}.

\section{Conclusions}

Measuring CP asymmetries in $B$ decays is important for several reasons:
\begin{itemize}
\item CP violation should be observed outside the $K$ system
($B^0\to J/\psi K_S$).
\item In many cases relative signs of asymmetries test the CKM picture (U-spin
related processes).
\item Certain asymmetries are predicted by CKM to be very small
($B_s\to J/\psi\phi$). Sizable asymmetries are signals of new physics.
\item Asymmetries in different processes test and overconstrain the CKM
parameters.
In certain cases phase determinations are theoretically clean ($B^0\to J/\psi
K_S,~B\to DK,~B_s\to D_s K$), some are difficult ($B^0\to\pi^0\pi^0$), and
others
still involve theoretical uncertainties due to rescattering and SU(3) breaking
effects ($B/B_s\to PP$). There are ways to measure and set bounds on these
corrections. A combined experimental and theoretical effort should resolve
these
uncertainties.
\item CP asymmetries will make us happy and our work interesting in the
next few years.
\end{itemize}

\section*{Acknowledgments}
I am indebted to David Atwood, Amol Dighe, Gad Eilam, David London, Dan
Pirjol,
Amarjit Soni, Daniel Wyler, and in particular to Jonathan Rosner
for very pleasant collaborations on these subjects.
This work was supported in part by the Israel Science Foundation founded by
the
Israel Academy of Sciences and Humanities, by the United States -- Israel
Binational Science Foundation under Research Grant Agreement 98-00237, and by
the fund for the promotion of research at the Technion.
\bigskip

\def \ajp#1#2#3{Am. J. Phys. {\bf#1}, #2 (#3)}
\def \apny#1#2#3{Ann. Phys. (N.Y.) {\bf#1}, #2 (#3)}
\def \app#1#2#3{Acta Phys. Polonica {\bf#1}, #2 (#3)}
\def \arnps#1#2#3{Ann. Rev. Nucl. Part. Sci. {\bf#1}, #2 (#3)}
\def \art{and references therein}
\def \cmts#1#2#3{Comments on Nucl. Part. Phys. {\bf#1}, #2 (#3)}
\def \cn{Collaboration}
\def \ite{{\it et al.}}
\def \cp89{{\it CP Violation,} edited by C. Jarlskog (World Scientific,
Singapore, 1989)}
\def \epjc#1#2#3{Eur.~Phys.~J.~C {\bf #1}, #2 (#3)}
\def \epl#1#2#3{Europhys.~Lett.~{\bf #1}, #2 (#3)}
\def \ib{{\it ibid.}~}
\def \ibj#1#2#3{~{\bf#1}, #2 (#3)}
\def \ijmpa#1#2#3{Int. J. Mod. Phys. A {\bf#1}, #2 (#3)}
\def \jpb#1#2#3{J.~Phys.~B~{\bf#1}, #2 (#3)}
\def \jhep#1#2#3{JHEP {\bf#1}, #2 (#3)}
\def \mpla#1#2#3{Mod. Phys. Lett. A {\bf#1}, #2 (#3)}
\def \nc#1#2#3{Nuovo Cim. {\bf#1}, #2 (#3)}
\def \np#1#2#3{Nucl. Phys. {\bf#1}, #2 (#3)}
\def \pisma#1#2#3#4{Pis'ma Zh. Eksp. Teor. Fiz. {\bf#1}, #2 (#3) [JETP Lett.
{\bf#1}, #4 (#3)]}
\def \pl#1#2#3{Phys. Lett. {\bf#1}, #2 (#3)}
\def \pla#1#2#3{Phys. Lett. A {\bf#1}, #2 (#3)}
\def \plb#1#2#3{Phys. Lett. B {\bf#1}, #2 (#3)}
\def \pr#1#2#3{Phys. Rev. {\bf#1}, #2 (#3)}
\def \prc#1#2#3{Phys. Rev. C {\bf#1}, #2 (#3)}
\def \prd#1#2#3{Phys. Rev. D {\bf#1}, #2 (#3)}
\def \prl#1#2#3{Phys. Rev. Lett. {\bf#1}, #2 (#3)}
\def \prp#1#2#3{Phys. Rep. {\bf#1}, #2 (#3)}
\def \ptp#1#2#3{Prog. Theor. Phys. {\bf#1}, #2 (#3)}
\def \ptwaw{Plenary talk, XXVIII International Conference on High Energy
Physics, Warsaw, July 25--31, 1996}
\def \rmp#1#2#3{Rev. Mod. Phys. {\bf#1}, #2 (#3)}
\def \rp#1{~~~~~\ldots\ldots{\rm rp~}{#1}~~~~~}
\def \stone{{\it $B$ Decays} (Revised 2nd Edition), edited by S. Stone
(World Scientific, Singapore, 1994)}
\def \yaf#1#2#3#4{Yad. Fiz. {\bf#1}, #2 (#3) [Sov. J. Nucl. Phys. {\bf #1},
#4 (#3)]}
\def \zhetf#1#2#3#4#5#6{Zh. Eksp. Teor. Fiz. {\bf #1}, #2 (#3) [Sov. Phys. -
JETP {\bf #4}, #5 (#6)]}
\def \zpc#1#2#3{Zeit. Phys. C {\bf#1}, #2 (#3)}
\def \zpd#1#2#3{Zeit. Phys. D {\bf#1}, #2 (#3)}
\def \PDG{Particle Data Group, D. E. Groom \ite, \epjc{15}{1}{2000}}


\end{document}